\documentclass[showpacs,twocolumn,aps]{revtex4}
\usepackage{amssymb}
\usepackage{amsmath}
\usepackage{graphicx}
\usepackage{lscape}
\usepackage{booktabs}
\begin{document}
\title{Exploring behaviors of partonic matter forming in early stage of $pp$ and 
       nucleus-nucleus collisions at ultra-relativistic energies}
\author{Ben-Hao Sa$^{1,2,3}$\footnote{sabh@ciae.ac.cn}, Dai-Mei Zhou$^{2}$,
Yu-Liang Yan$^{1}$, Bao-Guo Dong$^{1,4}$, Hai-Liang Ma$^{1}$, Xiao-Mei Li$^{1}$}
\address{1 China Institute of Atomic Energy, P.O. Box 275(18), Beijing 102413, China\\
2 Institute of Particle Physics, Huazhong Normal University, Wuhan 430079, China\\
3 CCAST (World Laboratory), P. O. Box 8730 Beijing 100080, China\\
4 Center of Theoretical Nuclear Physics, National Laboratory of Heavy Ion Collisions,
Lanzhou 730000,China}

%\date{\today}
\begin{abstract}
The parton yield, (pseudo)rapidity distribution, and transverse momentum distribution
in partonic matter assumed forming in the early stage of $pp$ and nucleus-nucleus
collisions at RHIC energy ($\sqrt{s_{{NN}}}$=200 GeV) and LHC energy ($\sqrt{s_{{NN}
}}$=5.5 TeV for nucleus-nucleus, 5.5 and/or 14 TeV for $pp$) are comparatively 
investigated with parton and hadron cascade model PACIAE. It turned out that the 
different parton and anti-parton spectra approach to be similar with reaction energy 
increasing from RHIC to LHC. We have argued that if the partonic matter forming 
in Au+Au collisions at RHIC energy is strongly interacting quark-gluon plasma, the 
one forming in Pb+Pb collisions at LHC energy might approach the real (free) 
quark-gluon plasma.
\end{abstract}

\maketitle

The coalescence (recombination) models \cite{biro,csiz,lin,hwo,frie,mol} have achieved
great successes in relativistic heavy-ion collisions. In these models, different to
the string fragmentation model, the relevant degrees of freedom are not ``free" quarks
but massive (dressed) quarks. The gluons are assumed to convert to quark pairs. It is
always assumed in these models that the spectrum of coalesced hadron is proportional
to the spectrum product of the coalescing partons and/or anti-partons. Here the parton 
refers to $d$, $u$, $s$, ... quarks, and gluons, anti-parton refers to $\bar d$, $\bar 
u$, $\bar s$, ... . The coalescence (recombination) models require the knowledge of 
parton $p_T$ and/or $\eta$ distributions as inputs \cite{ko}. Similarly, the 
perturbative quantum chromo-dynamics (pQCD) calculations \cite{wang,zhan} for hadron 
production in the relativistic elementary and/or nuclear collisions need the intrinsic 
transverse momentum distribution of parton in the nucleon as well.

Based on the further assumption of different quark and antiquark have same spectrum,
coalescence (recombination) models predicted that the elliptic flow parameter $v_2$
follows a quark-number ($n_q$) scaling in $v_2(p_T/n_q)$ vs. $p_T/n_q$ for most final
state hadrons in the intermediate $p_T$ region \cite{mol,mul,ko,hwo1}. Later, this
quark-number scaling prediction has been proved experimentally \cite{star2,huan}.
This $v_2$ quark-number scaling observation relates the spectrum of final state
hadron to the spectrum of initial state parton directly.

Recently, the concept of effective constituent quark has been introduced to connect
with the quark coalescence (recombination) picture \cite{huan,chen}. They assume that
the hadron's $p_T$ is composed of its effective constituent quark's $p_T/n$ (n is the
number of effective constituent quarks in hadron) and the different quark and
anti-quark have same $p_T$ distribution. Then they extracted the $p_T$ distribution 
of effective $u$ ($d$) and $s$ quarks from the ratio of $\Xi(p_T/3)/\phi(p_T/2)$
and $\Omega(p_T/3)/\phi(p_T/2)$, respectively, in Au+Au collisions at $\sqrt{s_
{{NN}}}$=200 GeV.

As mentioned in \cite{star1} recently that, ``... there has been an effort to fully
reconstructed jets in heavy-ion collisions in order to provide a direct measurement
of the partonic kinematics, independently of the fragmentation process (quenched or
unquenched)." In fact, experimentally extracting the partonic observables from final
state hadrons with reconstruction method have already become interesting 
\cite{star3,poch}. So it is important to investigate the properties (yield, $p_T$ 
distribution, and $\eta$ distribution) and explore their different behaviors in $pp$ 
and nucleus-nucleus collisions at RHIC energy ($\sqrt{s_{{NN}}}$=200 GeV) and LHC 
energy ($\sqrt{s_{{NN}}}$=5.5 TeV for nucleus-nucleus, 5.5 and/or 14 TeV for $pp$) 
theoretically. These studies might shed light on the way toward QGP.

In this paper the parton and hadron cascade model, PACIAE \cite{sa}, is used to
investigate systematically the yield, (pseudo)rapidity distribution, and transverse
momentum distribution of partons and anti-partons in the partonic matter assumed 
forming in the early stage of $pp$ and nucleus-nucleus collisions at RHIC and LHC 
energies. The discrimination, in the yields, $\eta$ distributions, and $p_T$ 
distributions, between different parton and anti-parton, RHIC and LHC energies, as 
well as $pp$ and nucleus-nucleus collisions is discussed. We have turned out that 
the different parton and anti-parton spectra approach to be similar with 
reaction energy increasing from RHIC to LHC. It is argued that if the partonic 
matter forming in the early stage of Au+Au collisions at RHIC energy is a strongly 
interacting QGP (sQGP) \cite{brah,phob,star,phen} the one forming in Pb+Pb 
collisions at LHC energy might approach to the real (free) QGP (fQGP).

PACIAE is a parton and hadron cascade model \cite{sa} based on PYTHIA \cite{soj2}. 
PYTHIA is a model for hadron-hadron collisions. The PACIAE model is composed of four 
stages of the parton initialization, parton evolution (rescattering), hadronization, 
and hadron evolution (rescattering).

A nucleon-nucleon ($NN$) collision in the PYTHIA (PACIAE) model is decomposed into
parton-parton collisions. The hard and soft parton-parton collisions are described,
respectively, by the lowest-leading-order (LO) pQCD parton-parton cross section
\cite{comb} and an empirical method. The semihard (between hard and soft) QCD $2
\rightarrow 2$ processes are involved as well. Because the initial- and final-state
QCD radiation are considered, the PYTHIA (PACIAE) model generates a multijet event 
for a $NN$ collision. This is followed by Lund and/or Independent Fragmentation model 
in the PYTHIA model, so one obtains a hadronic state for a $NN$ (hadron-hadron, $hh$)
collision. Since above fragmentation is switched-off in the PACIAE model, so one
obtains a multijet event (composed of quarks, anti-quarks, and gluons) instead.

A nucleus-nucleus collision in the PACIAE model is decomposed into $NN$ collisions
according to the collision geometry. The nucleons in a colliding nucleus are arranged
randomly in coordinate space according to the Woods-Saxon distribution (for radius
$r$) and 4$\pi$ isotropic distribution (orientation). We assume $p_x=p_y=0$ and $p_z$
equals the beam momentum for every colliding nucleon. Assuming straight line 
trajectory for nucleons we can calculate the collision time for each $NN$ collision 
pair provided that the closest approaching distance between two colliding nucleons is 
less than or equal to $\sqrt{\sigma_{\rm{tot}}^{NN}/\pi}$ ($\sigma_{\rm{tot}}^{NN}$ 
refers to the total cross section of $NN$ collision). If every $NN$ collision is 
performed with the method in previous paragraph until the collision pair is exhausted
, we obtain a initial partonic state for a nucleus-nucleus collision.

\begin{table}[htbp]
\caption{Composition of the partonic matter forming in $pp$ and 0-5\%
most central nucleus-nucleus collisions at RHIC and LHC energies.}
\begin{tabular}{ccccc}
\hline
      & \multicolumn{2}{c}{pp}&  Au+Au& Pb+Pb\\
\cmidrule[0.25pt](l{0.05cm}r{0.05cm}){2-3} \cmidrule[0.25pt](l{0.05cm}r{0.05cm}){4-4}
\cmidrule[0.25pt](l{0.05cm}r{0.05cm}){5-5}
Energy (GeV)& 200& 14000& 200& 5500\\
\cmidrule[0.25pt](l{0.05cm}r{0.05cm}){2-3} \cmidrule[0.25pt](l{0.05cm}r{0.05cm}){4-4}
\cmidrule[0.25pt](l{0.05cm}r{0.05cm}){5-5}
$d$     & 2.09 & 5.25& 330 & 244\\
$\bar d$& 0.237& 3.47& 27.3& 112\\
$u$     & 3.98 & 7.08& 294 & 226\\
$\bar u$& 0.234& 3.47& 27.5& 112\\
$s$     & 0.123& 1.19& 13.5& 39.6\\
$\bar s$& 0.106& 1.17& 12.7& 39.5\\
$g$     & 4.82 & 39.8& 516 & 1351\\
\hline
\end{tabular}
\label{mul}
\end{table}

\bigskip \bigskip
\begin{figure}[htbp]
\includegraphics[width=7.0cm,angle=0]{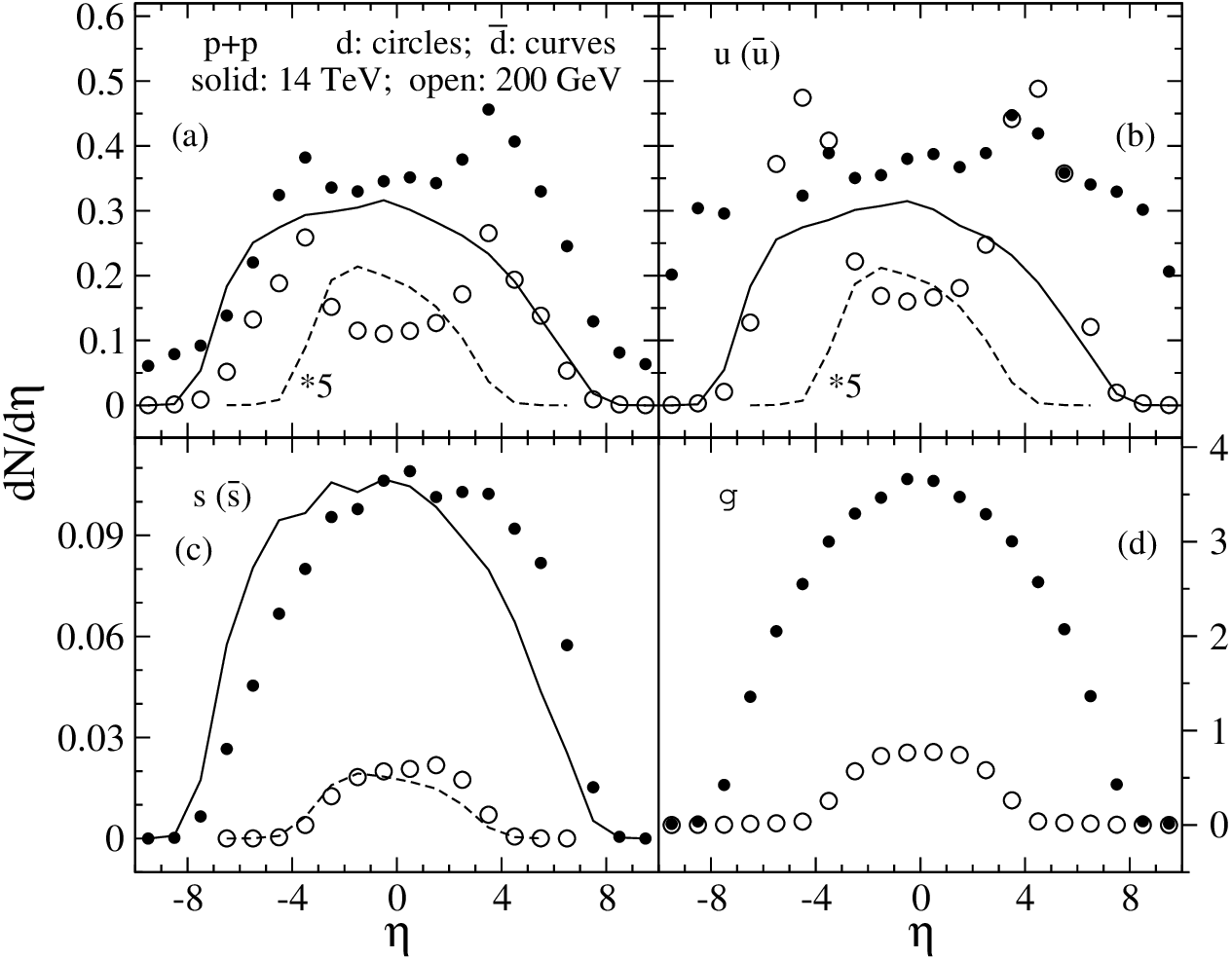}
\caption{Pseudorapidity distribution of parton and anti-parton in $pp$ collisions at 
$\sqrt{s}$=14000 and 200 GeV. The solid and open circles are for partons 
in $pp$ collisions at LHC and RHIC energies, respectively. Similarly the solid and 
dashed curves are for anti-partons. Panel (a), (b), (c), and (d) is for $d$ and $\bar 
d$, $u$ and $\bar u$, $s$ and $\bar s$, and $g$, respectively. The dashed curve has 
scaled by 5 in panels (a) and (b).} 
\label{ppeta}
\end{figure}

In the parton evolution (partonic rescattering) stage the $2\rightarrow2$ LO-pQCD
differential cross sections \cite{comb} are employed. For a subprocess $ij\rightarrow
kl$ the differential cross section reads
\begin{equation}
\frac{d\sigma_{ij\rightarrow
kl}}{d\hat{t}}=K\frac{\pi\alpha_s^2}{\hat{s}}\sum_{ij\rightarrow
kl},
\end{equation}
where the $K$ factor is introduced for the higher order corrections and the
nonperturbative correction as usual. Take the process $q_1q_2 \rightarrow q_1q_2$ as
an example one has
\begin{equation}
\sum_{q_1q_2\rightarrow
q_1q_2}=\frac{4}{9}\frac{\hat{s}^2+\hat{u}^2}{\hat{t}^2}.
\label{eq3}
\end{equation}
It can be regularized as
\begin{equation}
\sum_{q_1q_2\rightarrow
q_1q_2}=\frac{4}{9}\frac{\hat{s}^2+\hat{u}^2}{(\hat{t}-\mu^2)^2},
\end{equation}
by introducing the parton colour screen mass, $\mu$=0.63 GeV. In above equation
$\hat{s}$, $\hat{t}$, and $\hat{u}$ refer to the Mandelstam variables and $\alpha_s$=
0.47 stands for the running coupling constant. The total cross section of the parton
collision $i+j$ is then
\begin{equation}
\sigma_{ij}(\hat{s})=\sum_{k,l}\int_{-\hat{s}}^{0}d\hat{t}
\frac{d\sigma_{ij\to kl}}{d\hat{t}}.
\end{equation}
With the total and differential cross sections above the parton evolution
(rescattering) can be simulated by the Monte Carlo method until the parton-parton
collision is ceased (partonic freeze-out).

The hadronization follows parton evolution. In the PACIAE model partons can be 
hadronized by the string fragmentation scheme or the coalescence picture. As all 
the simulations are ended after parton rescattering in this paper, we do not describe 
the hadronization in detail but refer to Ref. \cite{sa}.

After hadronization we obtain a hadron list composed of the spatial and momentum
coordinates of all hadrons for a nucleus-nucleus collision. Similar to the  
parton rescattering above, one calculates the collision time for each $hh$ collision 
pair. Then each $hh$ collision is performed with usual two-body collision method 
\cite{sa1} until the collision pairs are exhausted (hadronic freeze-out).

As we aim at the physics rather than reproducing the experimental data, so in the 
calculations default values given in the PYTHIA model are adopted for model parameters 
except $K$=3 is assumed. The simulations are all ended at parton evolution 
(rescattering) stage.

\bigskip \bigskip
\begin{figure}[htbp]
\includegraphics[width=7.0cm,angle=0]{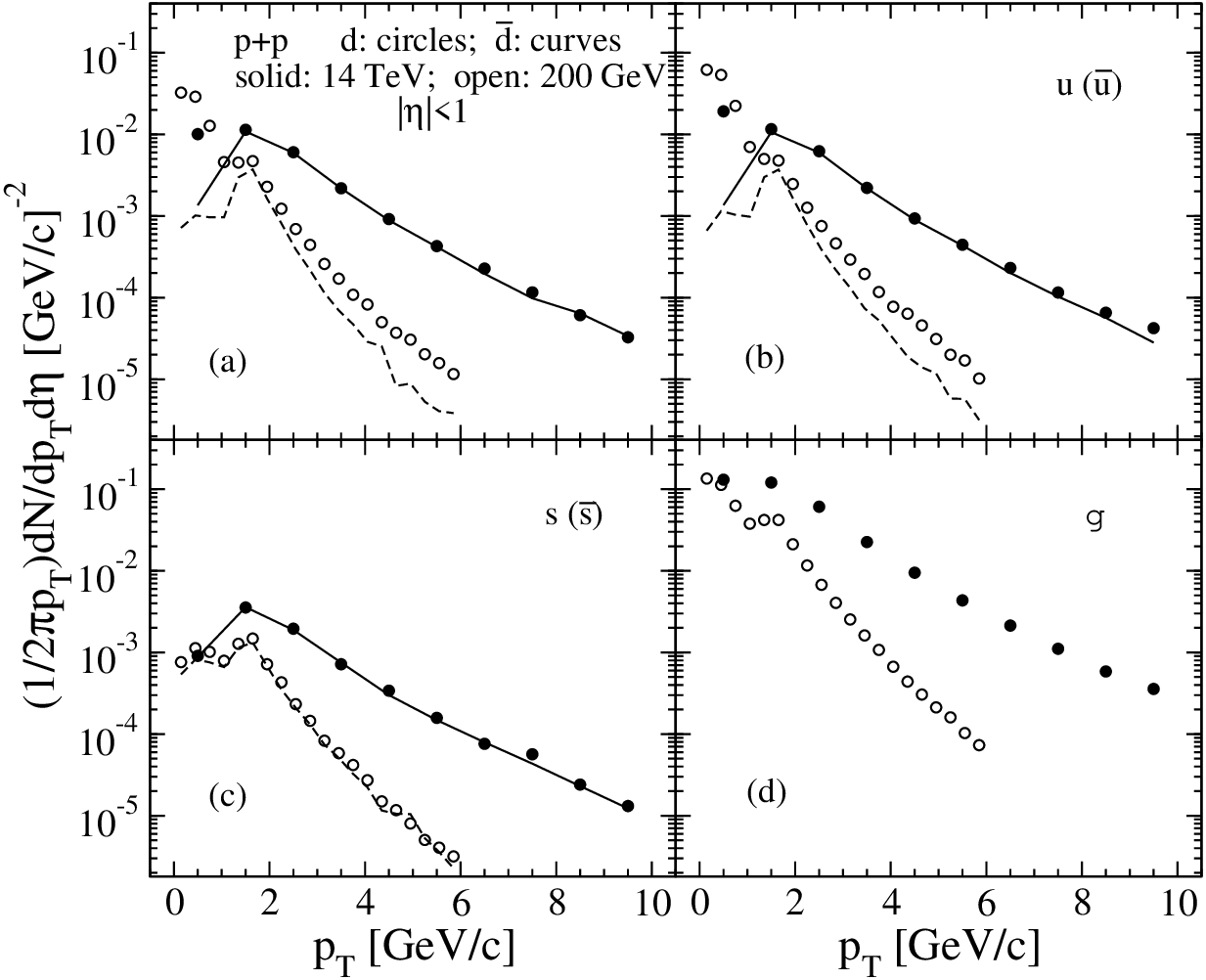}
\caption{Transverse momentum distribution of parton and anti-parton in $pp$ collisions 
at $\sqrt{s}$=14000 and 200 GeV. The solid and open circles are for 
partons in $pp$ collisions at LHC and RHIC energies, respectively. Similarly the solid 
and dashed curves are for anti-partons. Panel (a), (b), (c), and (d) is for $d$ and 
$\bar d$, $u$ and $\bar u$, $s$ and $\bar s$, and $g$, respectively.} 
\label{pppt}
\end{figure}

Four white papers \cite{brah,phob,star,phen} have explained that the partonic matter 
(quark gluon matter, QGM) formed in the early stage of Au+Au collisions at $\sqrt{s_
{{NN}}}$=200 GeV is sQGP. Suppose the QGM formed in the early stage of Pb+Pb collisions 
at $\sqrt{s_{{NN}}}$=5.5 TeV is a step toward fQGP. Then one could make conjectures for 
fQGP:
\begin{itemize}
\item According to the momentum fraction carried by the gluons in proton total momentum 
is 0.5 \cite{ff1} and inspired by the gluon saturation concept in the colour glass 
condensate model \cite{cgc}, we suppose the above fraction in Pb+Pb collision at 
LHC energy is lager than that in Au+Au collision at RHIC energy, because of the more 
gluon condensate in former reaction than in latter one. This fraction factor becomes 
larger than 0.5 with the QGM approaching fQGP.
\item In the Lund string fragmentation regime, the flavor selection of $u : d : s : c 
... \approx$ 1 : 1 : 0.3 : 10$^{-11}$ ... and $u\bar u : d\bar d : s\bar s, ... \approx 
$ 1 : 1 : 0.3 ... \cite{soj2} is introduced in hadron production based on the mass 
(dressed mass) effect. As this effect decreases with momentum (reaction energy
/temperature) increasing \cite{bhag}, it is reasonable to assume that a flavor 
balance of $u : d : s : c ... \approx$ 1 : 1 : $\gamma_s$ : $\gamma_c$ ... and $u\bar 
u : d\bar d : s\bar s, ... \approx$ 1 : 1 : $\gamma_s$ ... ($\gamma_s > $ 0.3 and 
$\gamma_c > $ 10$^{-11}$ for instance) is expected in fQGP.
\item It is well known the anti-particle to particle ratio approaches to one 
with energy (temperature) increasing, the ratio of anti-parton to parton is assumed 
approaching to one in fQGP.
\end{itemize}
The transport model results in this paper support above three conjectures.

In Tab.~\ref{mul} we give the parton chemical composition of partonic matter assumed
forming in the early stage of $pp$ and nucleus-nucleus collisions at RHIC and LHC 
energies. One sees in this table that:
\begin{itemize}
\item Assuming the sum of yields in each column is the total yield of partons in the
corresponding reaction. The fraction of gluons in total yield is about 0.420 and 0.422 
in $pp$ and Au+Au collisions at RHIC energy while those are about 0.648 and 0.636 in 
$pp$ and Pb+Pb collisions at LHC energy. Similarly the momentum fraction carried by 
gluons in total momentum is about 0.165 and 0.179 in $pp$ and Au+Au collisions at RHIC 
energy while those are about 0.420 and 0.415 in $pp$ and Pb+Pb collisions at LHC 
energy. These results indicate that the QGM forming in the early stage of Pb+Pb 
collisions at LHC energy is closer to fQGP than in Au+Au collisions at RHIC energy.
\item In both $pp$ and nucleus-nucleus collisions the yield of $\bar d$, $\bar u$, 
$s$, $\bar s$, and $g$ increases with reaction energy increasing from RHIC to LHC 
much stronger than $d$ and $u$. As the yield of $\bar d$, $\bar u$, $s$, $\bar 
s$, and $g$ in Pb+Pb collision at LHC energy is a few times larger than the 
corresponding one in Au+Au collision at RHIC energy, the yield of $d$ and $u$ in Pb+Pb 
collision is even somewhat less than that in Au+Au collision in order to meet with 
the above three conjectures.
\item $d$ and $u$ yields are about a magnitude larger than $\bar d$ and $\bar u$,
respectively, in $pp$ and Au+Au collisions at RHIC energy. But that yield difference 
drops to two dramatically in $pp$ and Pb+Pb collisions at LHC energy. This is 
consistent with the conjecture of the ratio of anti-parton to parton approaches one 
in fQGP.
\end{itemize}

\bigskip \bigskip 
\begin{figure}[htbp]
\includegraphics[width=7.0cm,angle=0]{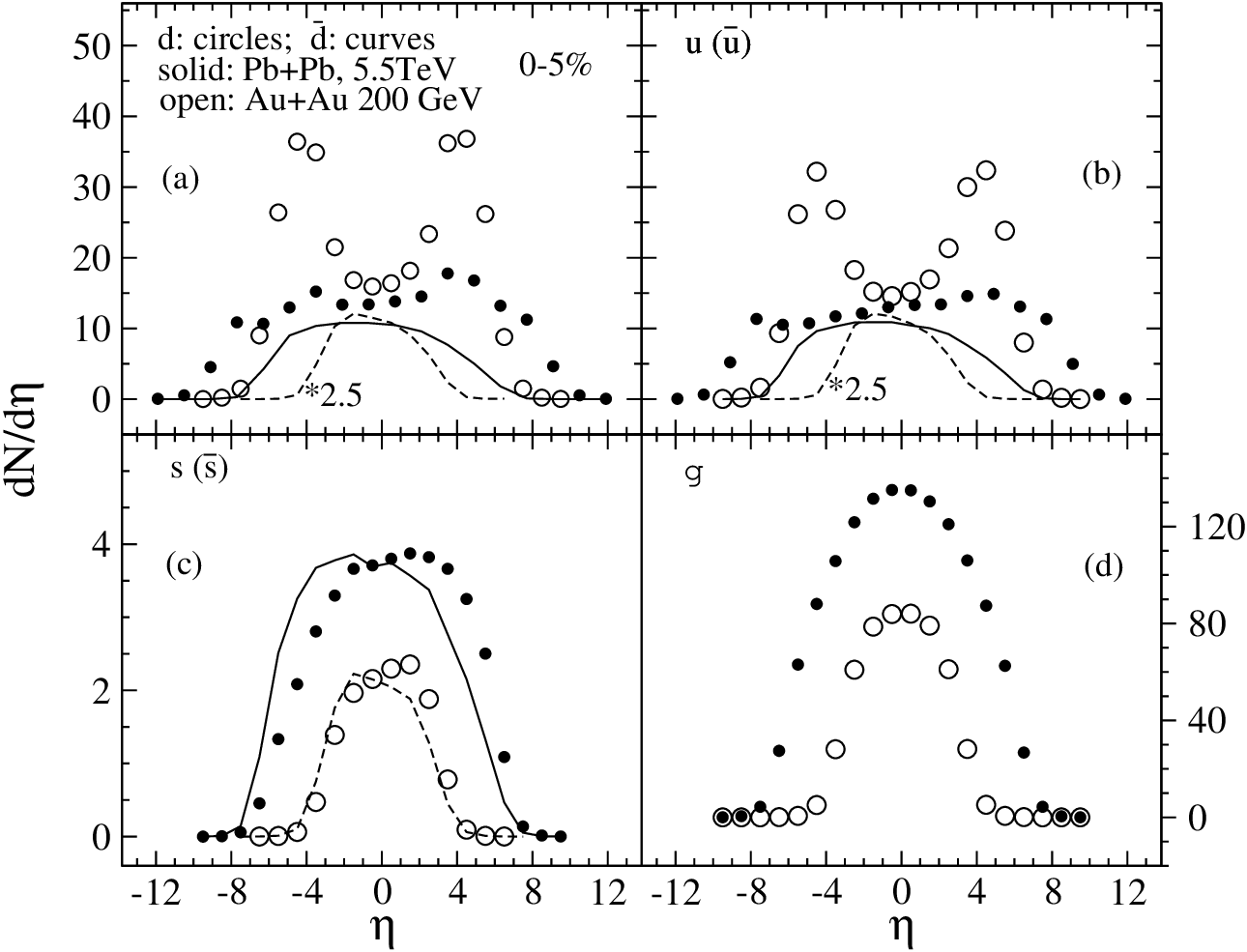}
\caption{The same as Fig.~\ref{ppeta} but for 0-5\% most central Pb+Pb and Au+Au 
collisions at $\sqrt{s_{{NN}}}$=5500 and 200 GeV instead of $pp$ collisions. The dashed 
curve has scaled by 2.5 in panels (a) and (b).} 
\label{peta}
\end{figure}

In Fig.~\ref{ppeta} we give the pseudorapidity distributions of partons and 
anti-partons in $pp$ collisions at $\sqrt s$=14000 and 200 GeV. One sees in panels (a) 
and (b) that the fragmentation peaks survive and midrapidity valley appears in the 
$\eta$ distribution of $d$ and $u$ quarks in $pp$ collisions at RHIC energy. But the 
fragmentation peaks approach to disappear and midrapidity valley fills up with energy 
increasing from RHIC to LHC. However the fragmentation peaks and midrapidity valley are 
not shown in the $\eta$ distribution of $\bar d$ and $\bar u$ quarks. The $\eta$ 
distribution of $u$ quark is higher (RHIC energy) or wider (LHC energy) than $d$ quark 
because proton is composed of $uud$. However the $\eta$ distribution of $\bar u$ quark 
is nearly the same as $\bar d$ in $pp$ collisions at both RHIC and LHC energies.

\bigskip \bigskip \bigskip
\begin{figure}[htbp]
\includegraphics[width=7.0cm,angle=0]{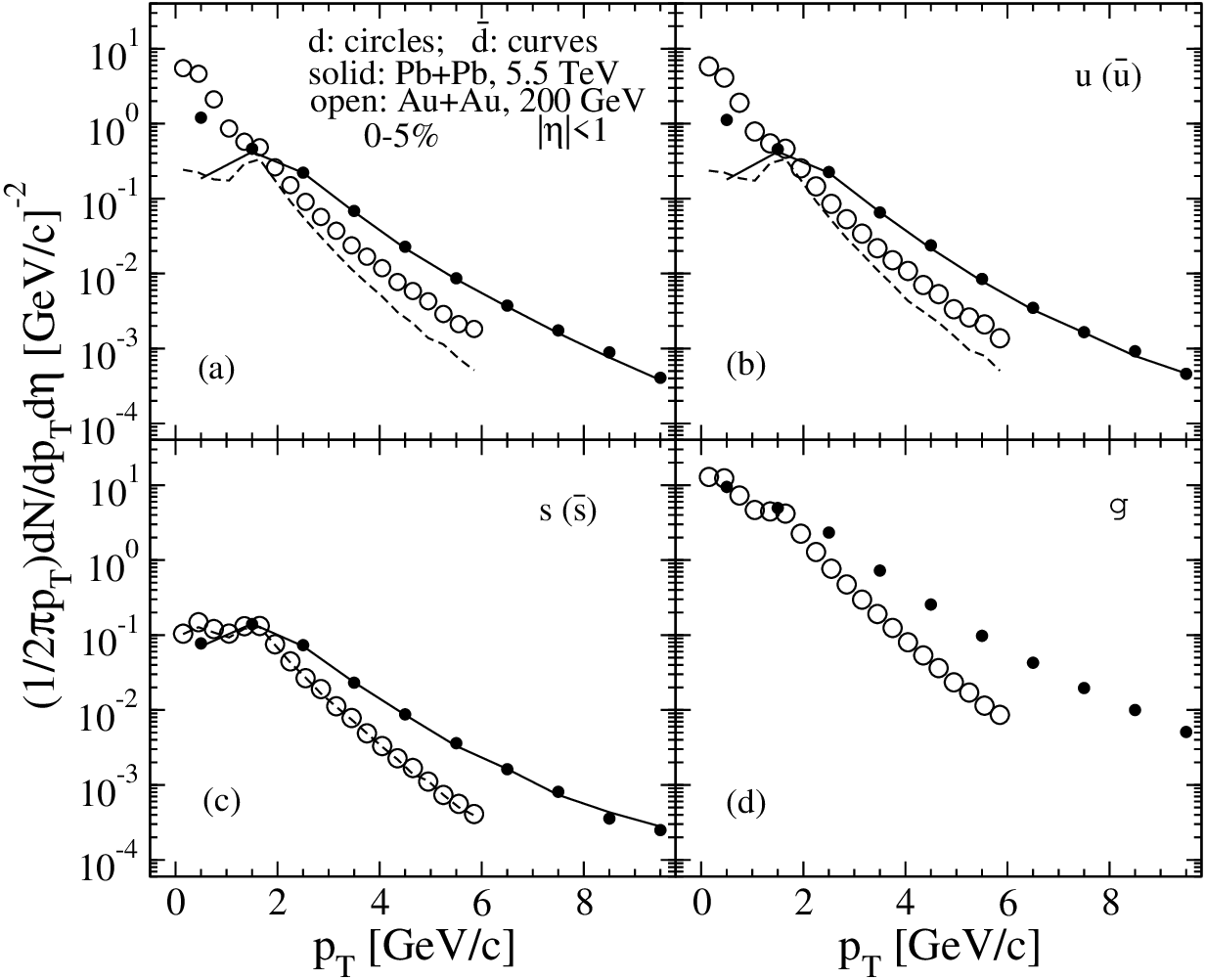}
\caption{The same as Fig.~\ref{pppt} but for 0-5\% most central Pb+Pb and Au+Au 
collisions at $\sqrt{s_{{NN}}}$=5500 and 200 GeV instead of $pp$ collisions.}
\label{ppt}
\end{figure}

Globally speaking, we know from Fig.~\ref{ppeta} that the $\eta$ distributions of 
partons and anti-partons in $pp$ collisions at LHC energy is more similar with 
each other than $pp$ collisions at RHIC energy. So it is reasonable to 
suppose that if QGM is not possible forming in the early stage of $pp$ collision at 
RHIC energy, it might be possible at LHC or higher energies. This idea is consistent 
with the knowledge of the nuclear system is able to deconfine by either highly 
heating or strongly compressing \cite{naga}.

The transverse momentum distributions of partons and anti-partons in $pp$ collisions 
at RHIC and LHC energies are given in Fig.~\ref{pppt}. We see in this figure that:
\begin{itemize}
\item The transverse momentum distribution of $d$ ($u$) quark is different from
$\bar d$ ($\bar u$), while the $p_T$ distribution of $s$ and $\bar s$ is nearly the 
same in $pp$ collisions at RHIC energy.
\item In $pp$ collisions at LHC energy the $p_T$ distribution of $d$ ($u$) is the 
same as $\bar d$ ($\bar u$) in the $p_T\geq$ 2 GeV/c region. But $s$ and $\bar s$ have 
the same $p_T$ distribution completely.
\item The slope parameter (apparent temperature) in the $p_T$ distribution of different
parton and anti-parton in $pp$ collisions at RHIC energy is larger (lower) than the 
corresponding one in $pp$ collisions at LHC energy.
\end{itemize}
These results indicate again that the different parton and anti-parton $p_T$ 
distributions approach similar with reaction energy increasing from RHIC to LHC. So 
one can not rule out the possibility of QGM forming in $pp$ collisions at LHC 
or higher energy.

Similar to Fig.~\ref{ppeta} we give the $\eta$ distributions of partons and
anti-partons in 0-5\% most central Pb+Pb and Au+Au collisions at $\sqrt{s_{{NN}}}$
=5500 and 200 GeV in Fig.~\ref{peta}, respectively. A discussion for Fig.~\ref{peta} 
parallel to Fig.~\ref{ppeta} can be drawn. It has only to emphasize again that the 
shapes of partons and anti-partons $\eta$ distributions are more similar with each 
other in Pb+Pb collisions at LHC energy than in Au+Au collisions at RHIC energy in 
globally speaking. Therefore if the partonic matter forming in the early stage of Au+Au 
collisions at RHIC energy is sQGP the one forming in Pb+Pb collisions at LHC energy 
might move forward to fQGP.

%\bigskip \bigskip \bigskip
\begin{figure}[htbp]
\includegraphics[width=7.0cm,angle=0]{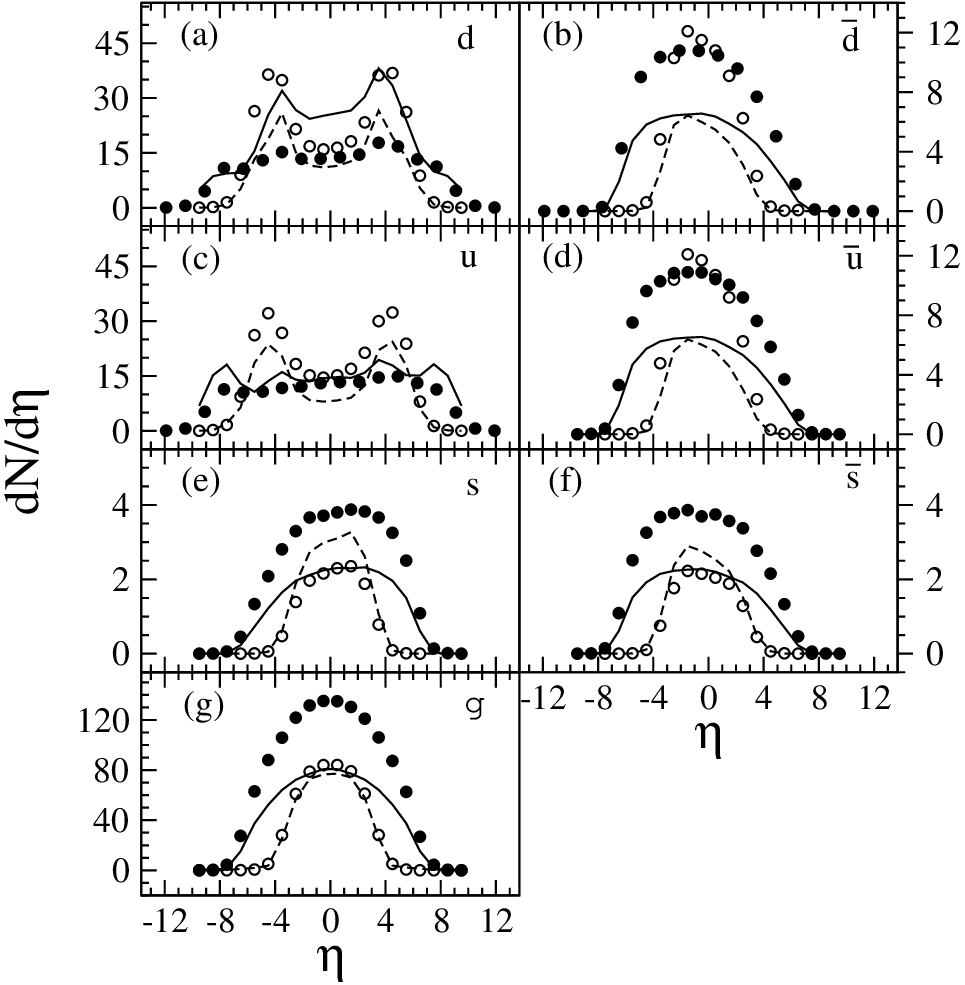}
\caption{Comparing pseudorapidity distributions among $pp$ collisions at $\sqrt s$= 
5500 and 200 GeV and 0-5\% most central Pb+Pb and Au+Au collisions at $\sqrt{s_{{NN}}}$
=5500 and 200 GeV. The panel (a), (b), (c), (d), (e), (f), and (g) is, respectively, for 
$d$, $\bar d$, $u$, $\bar u$, $s$, $\bar s$, and $g$. In any of these panels the solid 
and open circles are for the Pb+Pb and Au+Au collisions, respectively, and the solid and 
dashed curves for the $pp$ collisions. A scaling factor is introduced for $pp$ 
collisions, see text for the detail.}
\label{pneta}
\end{figure}

Figure ~\ref{ppt}, similar to the Fig.~\ref{pppt}, gives the $p_T$ distributions of
partons and anti-partons in 0-5\% most central Pb+Pb and Au+Au collisions at LHC 
and RHIC energies, respectively. In this figure one sees the same features as in 
Fig.~\ref{pppt}: The $p_T$ distribution of $d$ ($u$) quark is different from $\bar d$ 
($\bar u$), while the $p_T$ distribution of $s$ and $\bar s$ is the same in Au+Au 
collisions at RHIC energy. However in Pb+Pb collisions at LHC energy the different 
parton has the same $p_T$ distribution as corresponding anti-parton (for $d$ and $u$ 
quarks in the $p_T\geq$ 2 GeV/c region only). The slope parameters (apparent 
temperatures) in the $p_T$ distributions of different parton and anti-parton in Au+Au 
collisions at RHIC energy are larger (lower) than the corresponding ones in Pb+Pb 
collisions at LHC energy. Thus if the partonic matter forming in the early stage of 
Au+Au collisions at RHIC energy is sQGP the one forming in Pb+Pb collisions at LHC 
energy might approach fQGP.

\bigskip \bigskip \bigskip 
\begin{figure}[htbp]
\includegraphics[width=7.0cm,angle=0]{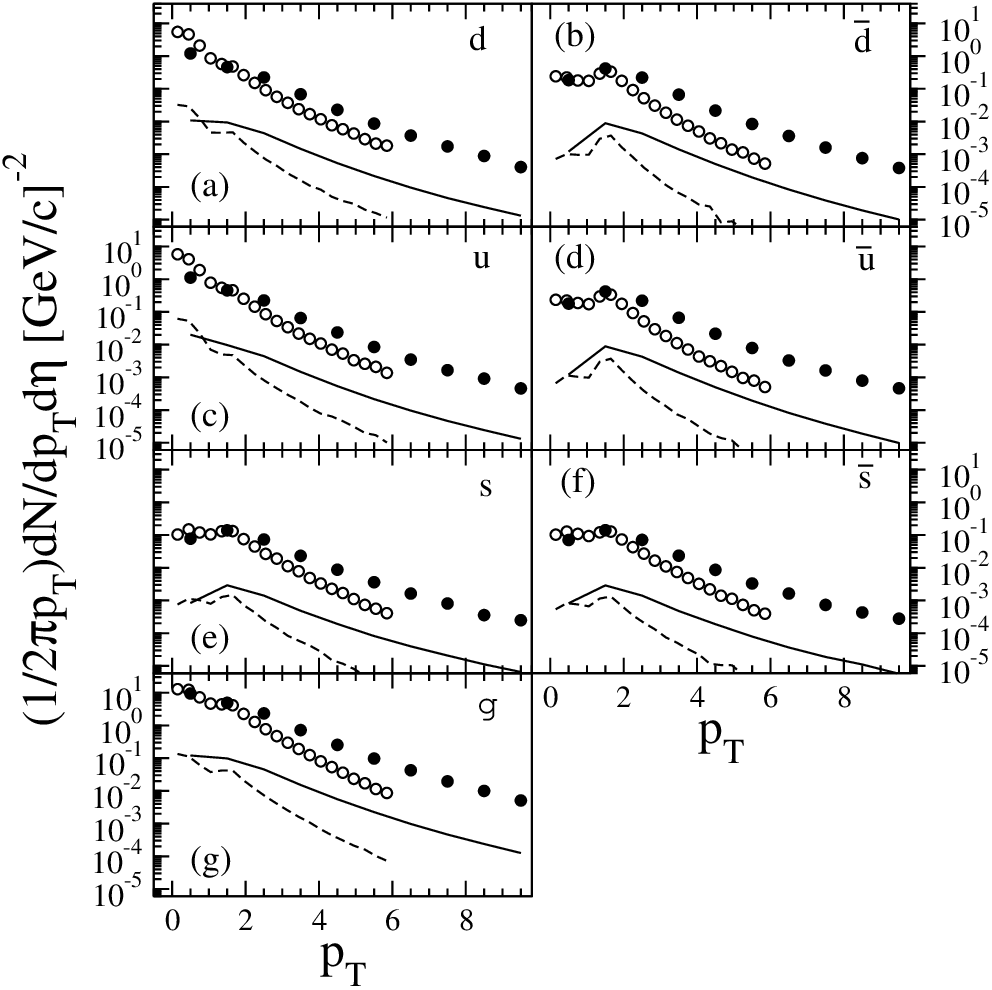}
\caption{The same as Fig.~\ref{pneta} but for transverse momentum distributions 
and no scaling factor is introduced.}
\label{pnpt}
\end{figure}

In Fig.~\ref{pneta} we compare the different parton and anti-parton $\eta$ 
distributions in $pp$ collisions at $\sqrt s$=5500 and 200 GeV with the corresponding 
ones in Pb+Pb and Au+Au collisions at the same cms energy, respectively. A scaling 
factor is introduced for $pp$ collisions at $\sqrt s$=5500 and 200 GeV. That is 100 and 
100 in panel (a), 30 and 150 in (b), 50 and 50 in (c), 30 and 150 in (d), 30 and 150 in 
(e), 30 and 150 in (f), and 30 and 100 in (g). We see in Fig.~\ref{pneta} that, in 
globally speaking the shapes of parton and anti-parton $\eta$ distributions in $pp$ 
collisions are similar to the corresponding ones in nucleus-nucleus collisions at 
the same cms energy. That is because in the PACIAE model the nucleus-nucleus 
collision is decomposed into $NN$ collisions and the $NN$ collision decomposes further 
into parton-parton collisions. Meanwhile, we consider only $2\rightarrow2$ 
parton-parton scattering processes and among these processes the elastic scattering 
is dominant. 

Similarly, one compares the different parton and anti-parton $p_T$ distributions in 
$pp$ collisions at $\sqrt s$=5500 and 200 GeV with the ones in Pb+Pb and Au+Au 
collisions at $\sqrt {s_{NN}}$=5500 and 200 GeV in Fig.~\ref{pnpt}, respectively. 
However, in Fig.~\ref{pnpt} we do not introduce any scaling factor. The same conclusion 
about the similarity discussed in the last paragraph may emerge if one examines the 
shapes of parton and anti-parton $p_T$ distributions in Fig.~\ref{pnpt}. Meanwhile, one 
sees in Fig.~\ref{pnpt} that the discrepancy in slope parameter (apparent temperature)  
between $pp$ and nucleus-nucleus collisions at the same cms energy decreases with 
cms energy increasing.
 
%\section{Conclusion}
In summary, we have used a parton and hadron cascade model, PACIAE, to investigate 
systematically the yield, (pseudo)rapidity distribution, and transverse momentum 
distribution of partons and anti-partons in partonic matter assumed forming in the 
early stage of $pp$ and nucleus-nucleus collisions at RHIC and LHC energies. The 
discrimination, in the yields, $\eta$ distributions, and $p_T$ distributions, between 
different parton and anti-parton, RHIC and LHC energies, as well as $pp$ and 
nucleus-nucleus collisions is discussed. We have turned out that the different parton 
and anti-parton approach to have similar spectrum with reaction energy increasing   
from RHIC to LHC. This is consistent with the fact of the (dressed) mass 
effect decreases with momentum (reaction energy/temperature) increasing. This result 
seems not supporting the assumption of different quark and anti-quark have the same 
$p_T$ distribution at RHIC energy. We have also argued that if the partonic matter 
forming in the early stage of Au+Au collisions at RHIC energy is sQGP, the one 
forming in the Pb+Pb collisions at LHC energy might approach fQGP.

The transport model results of this paper seem to support that the fQGP may properly 
have following physical features:
\begin{itemize}
\item Gluons carry more than half fraction in total momentum (energy).
\item Ratio of anti-parton to parton approaches one.
\item Flavor balance, such as \\
$u : d : s : c ... \approx$ 1 : 1 : $\gamma_s$ : $\gamma_c$ ...,\\
and $u\bar u : d\bar d : s\bar s, ... \approx$ 1 : 1 : $\gamma_s$ ...,\\
($\gamma_s\geq$ 0.3 and $\gamma_c\geq$ 10$^{-11}$ for instance) is
expected.
\end{itemize}
These physical features might shed light on the searching for fQGP.

Finally, the financial support from NSFC (10635020, 10605040, and 10705012) in China
is acknowledged

\end{document}